\documentclass[notoc]{JHEP3}
\usepackage{amsmath}
\usepackage{epsfig}
\usepackage{amssymb,amsfonts}

\def\eps{\varepsilon}
\def\Z{\mathbb{Z}}
\def\RP{\mathbb{RP}}
\def\O{{\cal O}}
\def\N{{\cal N}}
\def\A{{\cal A}}
\def\K{{\cal K}}
\def\M{{\cal M}}
\def\>{\rangle}
\def\I{\mathbb{I}}

\title{D-branes and orientifolds of SO(3)}
\author{N. Couchoud \\ 
Laboratoire de Physique Th\'eorique et Hautes Energies%
\thanks{Unit\'e mixte du CNRS et de l'Universit\'e de Paris VI et Paris VII,
UMR 7589.} \\
Universit\'e Pierre et Marie Curie,  Paris VI \\
4 place Jussieu, 75252 Paris CEDEX 05, France \\
\email{couchoud@lpthe.jussieu.fr}\\
\\
Laboratoire de Physique Th\'eorique de l'\'Ecole Normale Sup\'erieure%
\thanks{Unit\'e mixte du CNRS et de l'\'Ecole Normale Sup\'erieure,
UMR 8549.} \\
24 rue Lhomond, 75231 Paris CEDEX 05, France \\
\email{couchoud@lpt.ens.fr}
}

\abstract{
We study branes and orientifolds on the group manifold of $SO(3)$. We
consider particularly the case of the equatorial branes, which are
projective planes. We show that a Dirac-Born-Infeld action can be
defined on them, although they are not orientable. We find that there
are two orientifold projections with the same spacetime action, which
differ by their action on equatorial branes.
}

\keywords{D-branes, Conformal Field Models in String Theory}

\preprint{ LPTHE-02-03 \\ LPTENS-02/05 \\ \hepth{0201089} }

\begin{document}
\section{Introduction}
The purpose of this note is to study branes and orientifolds on the
group manifold of $SO(3)$ in a geometrical way. The group $SO(3)$ is
the quotient of $SU(2)$ by its center $\Z_2$, whose non-trivial
element acts as $g \to -g$, which is geometrically the antipodal map;
thus the geometry of $SO(3)$ is $\RP^3$.

Strings on $SO(3)$ are described by a $SU(2)$ WZW model with $k$ even.
Their torus amplitude, derived in \cite{GW}, is given by the
$D_{k/2+2}$ modular invariant. The spectrum of oriented closed strings
thus contains non-twisted states, whose left and right isospins $j$
and $\bar{j}$ are equal and integer, and twisted states with
$\bar{j}=k/2-j$, where $j$ is integer if $k$ is a multiple of 4, and
half-integer otherwise. In particular, there always exists a twisted
state with $j=\bar{j}=k/4$ and this is the only one with $j=\bar{j}$.

First, we consider D-branes on $SO(3)$. They were studied in various
papers, in a purely algebraic way \cite{Sag,SS} or in a more geometric
manner \cite{FFFS,DMS,MSS}. 

Like in any WZW model, the maximally symmetric branes (i.e.\ those
that preserve the Kac-Moody symmetry) are localized on conjugacy
classes of the group. What is particularly interesting here is the
conjugacy class located at the ``equator'' (the rotations of angle
$\pi$), which is a projective plane. This gives rise to two sorts of
$\RP^2$ branes, which differ only by their coupling to twisted closed
strings; they can be combined in pairs into wrapped spherical branes.

We also show that a Dirac-Born-Infeld action can be defined on the
$\RP^2$ brane, although it is not orientable. Then we compute the
spectrum of its small fluctuations like in \cite{BDS}, and we find
that it corresponds to the spectrum of the open strings, as expected.

Next, we consider the orientifolds. They were studied purely
algebraically in \cite{Sag,SS}, and only recently in a geometric way:
\cite{Bru,HSS} study them on general WZW models, and \cite{BCW} studies
in great detail the case of the three-sphere, including the relations
between branes and orientifolds. Here we will study the $SO(3)$ case
in a similar manner.

We find that there are two orientifold projections with the same
orientifold fixed points. These projections essentially differ by
their actions on equatorial branes: in one case, the projection acts
separately on the two sorts of $\RP^2$ branes; in the other case, only
spherical branes exist and the projection acts on them as a whole.

All the results found here geometrically are consistent with the
previous algebraic results.

\section{The D-branes and the annulus amplitude}
\subsection{Geometrical description of the D-branes}
The simplest way of studying D-branes on $SO(3)$ is by describing them
as branes of $SU(2)$ with the $\Z_2$ antipodal identification.

On $SU(2)$, which is simply connected, maximally symmetric branes are
Cardy states \cite{Car}; geometrically, the Cardy state associated
with the representation of isospin $j$ is a sphere, whose equation in
the usual spherical coordinates is $\psi = (2j+1)\pi/(k+2)$
\cite{AS,FFFS}. Now the antipodal identification acts as
\begin{equation}
\label{Z2}
\psi \to \pi-\psi\ , \quad \theta \to \pi-\theta\ ,
\quad \phi \to \phi+\pi\ ,
\end{equation}
which transforms the Cardy state of isospin $j$ into the Cardy state
of $k/2-j$.

So, for $j < k/4$, the two distinct branes corresponding to the Cardy
states of isospin $j$ and $k/2-j$ of $SU(2)$ are identified; thus, one
obtains spherical branes, which will be called ``branes of type $j$''.

On the other hand, on the ``equator'', which is a projective plane
because of the $\Z_2$ identification, one can have two sorts of
branes:
\begin{itemize}
\item by identifying a point on an equatorial brane of $S^3$ with the
opposite point on {\em another} equatorial brane, one obtains a
spherical brane wrapped on the equator (i.e.\ on any distinct point of
the equator there are two points of the brane).
\item by identifying a point on an equatorial brane of $S^3$ with the
opposite point on the {\em same} brane, one obtains a projective
plane brane.
\end{itemize}

More precisely, to derive the spectrum and the gauge group concerning
the equatorial branes, we will use the following description: we have
$N$ equatorial spheres on $S^3$, and the $\Z_2$ identification acts as
an involutive permutation of them. Now a boudary state can be an
arbitrary linear combination of branes, so the space of boundary
states is a $N$-dimensional vector space and $\Z_2$ acts on it as a
$N \times N$ unitary matrix $Z$. $Z^2=1$, so it is diagonalizable with
$n_+$ eigenvalues $+1$ and $n_-$ eigenvalues $-1$, with $n_++n_-=N$,
i.e.\ our branes can be described as $n_+$ $\RP^2$ branes with
sign~$+$, and $n_-$ $\RP^2$ branes with sign~$-$. Thus the equatorial
brane of the three-sphere gives rise to two sorts of branes (this was
observed in \cite{FFFS}). An interpretation of these signs will be
given in the next section.

Now spherical equatorial branes correspond to $\displaystyle 
Z= \left( \begin{array}{cc} 0 & 1 \\ 1 & 0 \\ \end{array} \right)$,
which has one $+1$ and one $-1$ eigenvalue, so a spherical equatorial brane
is a combination of two $\RP^2$ branes of opposite signs.

\subsection{The annulus amplitude}
Strings between a D-brane of type $r$ and one of type $s$ correspond
in the 3-sphere to strings between branes of isospins $r$ and $s$ on
one hand, and $k/2-r$ and $s$ on the other hand (the two other
possibilities are identified with these ones). So the corresponding
annulus reads:
\begin{equation}
\A_{rs} = \sum_{l=0}^{k/2}\;
	  (N_{rs}{}^l + N_{k/2-r,s}{}^l)\; \chi_l(\sqrt{q})
\end{equation}
where $q=e^{-2\pi t}$, $\chi_l$ are the Kac-Moody characters, and
$N_{rs}{}^l$ are the fusion coefficients of the $SU(2)$ Kac-Moody
algebra, whose non-zero values are
\begin{equation}
N_{rs}{}^l = 1,
\text{ for } l= |r-s|, \; |r-s|+1 , \;\cdots, \; {\rm min}(r+s, k-r-s)\ .
\end{equation}
($l$ can be integer or half-integer except when stated otherwise.)

In order to extract the D-brane couplings to the closed-string modes
we must rewrite these amplitudes in the dual `transverse' channel.
This is achieved by changing variables
\begin{equation}
q= e^{- 2\pi  t} \rightarrow {\tilde q} = e^{-2\pi/ t}\ , 
\end{equation}
and using  the  modular property  of the characters
\begin{equation}
\chi_l(\sqrt{q}) = S_l{}^j\;  \chi_j ({\tilde q}^2)\ .
\end{equation}
Consistency requires the final result to be of the general form
\begin{equation}
\A_{rs} = \sum_{j=0}^{k/2}\; D^j_r D^j_s \; \chi_j({\tilde q}^2)\ ,
\end{equation}
where $D^j_s$ is the coupling of a D-brane of type $s$ to closed
strings in the $(j,j)$ representation of the Kac-Moody algebra.

The modular-transformation matrix for the WZW model reads:
\begin{equation}
\label{S}
S_l{}^j = \sqrt{\frac{2}{k+2}}\;
          \sin\left(\frac{(2l+1)(2j+1)\pi}{k+2}\right)\ .
\end{equation}
One then finds
\begin{equation}
D^j_s = E_{2j}\; \sqrt{2}\;
        \sin\left(\frac{(2j+1)(2s+1)\pi}{k+2}\right)\;
        \sqrt{\N_j}
\end{equation}
where $E_n$ is a projector onto even integers (1 if $n$ is even, 0
otherwise), and ${\cal N}_j$ is a normalization such that
\begin{equation}
(\N_j)^{-1} \equiv \sqrt{\frac{k+2}{2}}\;\;
                         \sin\left(\frac{(2j+1)\pi}{k+2}\right)\ .
\end{equation}
Thus, up to a factor $\sqrt{2}$ (and a $E_{2j}$ because $(j,j)$
untwisted strings with j half-integer do not exist), these couplings
are the same as in the $SU(2)$ case \cite{BCW}.

Note that, for topological reasons, the spherical branes cannot couple
to twisted closed strings, so that these are couplings to untwisted
closed strings only. Moreover, as spherical equatorial branes are
quite similar to other spherical branes, they can be considered in
this calculation as branes of type $k/4$, so their couplings are equal
to $D^j_{k/4}$.

For an open string stretching between an $\RP^2$ D-brane and one of
type $s$ the annulus reads:
\begin{equation}
\A_{Rs} = \sum_{l=0}^{k/2} N_{k/4,s}{}^l\; \chi_l(\sqrt{q})
\end{equation}
(with $R$ standing for $\RP^2$). Note that this is independent of the sign
of the $\RP^2$ brane. One then finds
\begin{equation}
\label{DjR}
D^j_R = \frac{1}{2} D^j_{k/4}
\end{equation}
which is not surprising, since spherical equatorial branes are a combination
of two $\RP^2$ branes. Note that this amplitude does not contain
any information about couplings of the $\RP^2$ brane with
twisted strings.

Now we consider strings stretching between two $\RP^2$ branes of
definite sign, which correspond on the 3-sphere to strings between
equatorial branes. As such states are eigenvalues of the action of
$\Z_2$, which we will also call $Z$, one half of the states are
projected out. One has
\begin{equation}
\label{Z}
Z|N,l,ij\> = (-)^l s_i s_j|N,l,ij\>
	     \text{ for } l=0,\; 1,\; \cdots,\; k/2\ .
\end{equation}
where $N$ is the excitation number of the state, $l$ the spin of the
SO(3) representation under which it transforms, $i$ and $j$ the
indices of the two branes, $s_i$ and $s_j$ their signs. The range of
$l$ corresponds to the non-zero values of $N_{k/4,k/4}{}^l$ and the
sign $(-)^l$ is the parity of the $l$th spherical harmonic. So we have
only strings of even isospin between branes of the same sign, and of
odd isospin between branes of opposite sign. Thus, the corresponding
annulus amplitude reads
\begin{equation}
\A_{RR} = \frac{1}{2} \sum_{\substack{l=0 \\ l\rm\ integer}}^{k/2}
	              \chi_l(\sqrt{q})
          \pm \frac{1}{2} \sum_{\substack{l=0 \\ l\rm\ integer}}^{k/2}\;
			  (-)^l \chi_l(\sqrt{q})\ .
\end{equation}
where $\pm$ is the product of both signs.

As the first term in the latter sum contains all the couplings of the
$\RP^2$ brane with untwisted strings, the second term contains only the
couplings with twisted strings. After a modular transformation, one
finds that the brane couples only with the twisted state with
$(j,\bar{j})=(k/4,k/4)$, and the coupling is
\begin{equation}
D^{\rm twisted}_R = \pm \frac{1}{2}\sqrt{\frac{k+2}{2}}\;
\end{equation}
where $\pm$ is the sign of the brane. So we see that the sign of a
brane can be interpreted, up to a numerical factor, as its charge
under the twisted state with $(j,\bar{j})=(k/4,k/4)$ (this was already
noticed in \cite{DMS}). Note that a spherical equatorial brane has
thus a vanishing charge under this twisted state, which is normal
since its topology does not allow any coupling between them.

Now let us compute the gauge symmetry group. As photons are strings
with both endpoints on branes of the same type, and the same twisted
charge for $\RP^2$ branes, the gauge group is $\prod_{j<k/4} U(n_j)
\times U(n_+) \times U(n_-)$. In particular, for $s$ spherical
equatorial branes and $r$ $\RP^2$ branes of the same charge (which is
the general case since $\RP^2$ branes of opposite charges can be
combined into spherical branes), the gauge group is $U(s+r) \times U(s)$.

Note that it is not clear whether or not a configuration with several
branes is stable; for example, in the $SU(2)$ case, it was shown (see
for example \cite{ARS}) that such a configuration can decay into a
single brane. We will not discuss stability considerations in this
paper.

\subsection{Semi-classical considerations}
Some of the results found here can be derived from a semi-classical
analysis, similar to what was done in \cite{BDS} for the 3-sphere.

In the case of spherical branes, we can apply the general results of
\cite{BRS}, and we find that the energy of the brane and the
fluctuations are essentially the same than on the 3-sphere. For the
spherical equatorial brane, the spectrum of small fluctuations
contains only half of the light open strings because this brane is
wrapped on a projective plane; an interpretation in terms of a
non-commutative geometry is given in \cite{MSS}.

In the case of equatorial $\RP^2$ branes, one may think that the
non-orientability of the brane makes the definition of a Born-Infeld
action impossible. In fact, this is not the case: to define the
integration of a real function on $\RP^2$, one just needs a measure;
since the measure $d\theta\,d\phi$ on $S^2$ is $\Z_2$-invariant, one
obtains a well-defined measure for $\RP^2$. The question is then to
check whether the integrand, i.e.\ the Born-Infeld Lagrangian, is
invariant. Without fluctuations, and taking fluxes $F=0$ and $B=0$ (a
two-form cannot have a non-vanishing constant flux on a non-orientable
surface), the Lagrangian just contains the metric, so the invariance
is trivial, making the Born-Infeld action well-defined. Thus the
energy of the $\RP^2$ brane is simply one half of the energy for an
equatorial spherical brane, which is not surprising.

Now consider small fluctuations of this brane. We write them like in
\cite{BDS}, parametrizing the worldvolume by $(t, \theta, \phi)$:
\begin{equation}
\psi = \frac{\pi}{2}+\delta\, , \quad
A_{\theta} = \frac{k}{2\pi}\alpha_{\theta} \quad 
\text{and } A_{\phi} = \frac{k}{2\pi}\alpha_{\phi}\ .
\end{equation}
Then the computation of the linearized equation of motion is
straightforward, and one finds
\begin{equation}
\frac{d^2}{dt^2}
\left( \begin{array}{c} \delta \\ f/\sin\theta \\ \end{array} \right)
= -\frac{1}{k\alpha'}
\left(
 \begin{array}{cc} \square+2 & \;2 \\ 2\,\square & \;\square \\ \end{array}
\right)
\left( \begin{array}{c} \delta \\ f/\sin\theta \\ \end{array} \right)
\end{equation}
with $f = \partial_{\theta}\alpha_{\phi} - \partial_{\phi}\alpha_{\theta}$.

Now the solutions we are looking for must be invariant under the
transformations \eqref{Z2}, i.e.\ one must have
\begin{eqnarray}
\delta(\pi-\theta, \phi+\pi) & = & -\delta(\theta, \phi) \\
A_{\theta}(\pi-\theta, \phi+\pi) & = & -A_{\theta}(\theta, \phi) \\
A_{\phi}(\pi-\theta, \phi+\pi) & = & A_{\phi}(\theta, \phi).
\end{eqnarray}
This implies that $\delta$ and $f/\sin\theta$ are odd functions, so
they contain only odd spherical harmonics. Finally, after some more
computations, the spectrum of quadratic fluctuations is
\begin{equation}
m^2 = \frac{l(l+1)}{k\alpha'} \quad\text{in reps. } (l-1) \oplus (l+1)
\text{ with $l$ even.}
\end{equation}
Thus the DBI spectrum is the same than the exact CFT result, except
for the fact that the former does not have any level-dependent
truncation.

\boldmath
\section{Orientifolds on $SO(3)$ and the Klein bottle amplitude}
\unboldmath

As explained in \cite{BCW}, the orientifold action is an orientation
reversal $\Omega$ on the world-sheet (i.e.\ $\Omega : z \to \bar z$),
which must be combined with an orientation-flipping isometry $h$ of
the target space to preserve the Wess-Zumino-Witten term in the
action. Such an isometry can be described as an isometry of $S^3$ with
$h^2 = \pm 1$. As the isometries with $h^2 = -1$ preserve the
orientation, the allowed transformations are the same than on the
three-sphere, i.e.\ $g \to \pm g^{-1}$. As these two possible
isometries are identified through $\Z_2$, we are left with only one,
whose fixed points are an $\O0$ at the pole (the two poles of $S^3$
are identified) and a projective plane $\O2$ at the ``equator''.

Now we can derive the Klein bottle amplitude. To do so, we have to
know how the various states transform, more precisely those with
$j=\bar j$ since only those contribute. The untwisted states are the
same than on the three-sphere with the constraint that $j$ must be an
integer, so they are invariant under $\Omega h$. The twisted states
with $j=\bar j=k/4$ also contribute, and since we do not know a priori
how they transform, we consider both possibilities. Thus, the Klein
bottle amplitude reads
\begin{equation}
\K = \sum_{\substack{l=0 \\ l\rm\ integer}}^{k/2}
     \chi_l(q^2) + \zeta \chi_{k/4}(q^2)
\quad\text{with}\quad \zeta = \pm 1.
\end{equation}

After a modular transformation with the $S$ matrix (eq. \eqref{S}),
the amplitude reads
\begin{equation}
\K = \sum_{j=0}^{k/2}(C^j)^2\chi_j(\sqrt{\tilde q\,})
\end{equation}
where $C^j$ is the coupling of the orientifold with the untwisted%
\footnote{A priori, one may expect that there be also a nonvanishing
coupling with twisted strings. This is inconsistent with the M\"obius
amplitudes}
closed strings. These `crosscap coefficients' are
\begin{equation}
C^j = \eps_j E_{2j} \sqrt{\N_j} 
      \left( \sin \left( \frac{(2j+1)\pi}{2k+4} \right) + 
        \zeta\; (-)^j \cos \left( \frac{(2j+1)\pi}{2k+4} \right) \right).
\end{equation}
Notice that this coefficient is similar to the sum of the coefficients
for the two sorts of orientifold projections of the three-sphere,
which meshes nicely with the fact that we have an $\O0$ and an $\O2$
together.

\section{Unoriented open strings and the M\"obius amplitude}
\subsection{Unoriented open strings}
Unoriented open strings are open strings which are invariant under
$\Omega h$, which transforms a brane into a brane of the same type
(but not necessarily of the same twisted charge in the case of $\RP^2$
branes) and exchanges the ends of the string. If a string has its ends
on branes of different types, a linear combination of that string and
its image by $\Omega h$ will remain. So we are interested in how
$\Omega h$ acts on the open strings between branes of the same type.

Strings between non-equatorial branes of type $r$ on $SO(3)$ correspond
to two inequivalent sorts of strings on $SU(2)$: strings with both ends
on a brane of type $r$, and strings with one end on a brane of type
$r$ and the other on a brane of type $k/2-r$. The action of $\Omega h$
on these strings has been derived in \cite{BCW}; its eigenvalue for a
string of isospin $l$ is proportional to $(-)^l$ in the former case,
and independant of $l$ in the latter case. The gauge group for $n_r$
branes is then $SO(n_r)$ or $USp(n_r)$, as usual.

In the case of equatorial $\RP^2$ branes, things are not so simple,
since branes of opposite charges may be mixed by the orientifold. The
most general action of the orientifold on the states is \cite{GP}
\begin{equation}
\Omega h|N,l,ij\> = (-)^{N+1}\gamma_{ii'}\gamma_{jj'}^*|N,l,j'i'\>
\end{equation}
where $\gamma$ is a unitary, symmetric or antisymmetric, matrix. Now,
on $SO(3)$, states must be invariant under the $Z$ transformation
\eqref{Z}, so we impose that $\Omega h$ acting on a $Z$-invariant
state results in a $Z$-invariant state. This implies that $\gamma$
must either commute or anticommute with $Z$.

If $\gamma$ commutes with $Z$, $+$ and $-$ branes are not mixed by the
orientifold, and the gauge group is $SO(n_+) \times SO(n_-)$ or
$USp(n_+) \times USp(n_-)$.

If $\gamma$ anticommutes with $Z$, which is possible only if
$n_+ = n_- \equiv n$, then, through changes of basis in the $+$ and
$-$ branes separately (i.e.\ that do not mix $+$ and $-$ branes),
$\gamma$ can be put into the form
\begin{equation}
\gamma = \left(\begin{array}{cr} 0 & \pm\I \\ \I & 0 \end{array}\right)\ ,
\end{equation}
where the first rows and columns of the matrix correspond to $+$
branes. So the effect of $\Omega h$ on the Chan-Paton coefficients is
\begin{equation}
\begin{array}{lrrl}
|++\> & \leftrightarrow &      |--\> \\
|+-\> & \to             & \pm\,|+-\> \\
|-+\> & \to             & \pm\,|-+\> & .
\end{array}
\end{equation}
Before orientifolding, one had one $U(n)$ group corresponding to
$|++\>$ states, and another for the $|--\>$ states. As they are
identified through the orientifold, one is left with a $U(n)$ gauge
group. Notice that $+$ and $-$ $\RP^2$ branes cannot exist separately
here, and that the $\gamma$ matrix defines pairs of branes of opposite
charges, so that it is better to think of them as $n$ spherical
equatorial branes. The $U(n)$ symmetry group is then interpreted as
arbitrary changes of basis in the space of these branes.

\subsection{The M\"obius amplitude}
As explained clearly in \cite{SS}, the M\"obius amplitude is given by a
linear combination of $\chi_l(-\sqrt{q})$ in the direct channel, and
$\chi_j(-\sqrt{\tilde{q}})$ in the transverse channel. Thus, it is
convenient to work in the basis of real characters
\begin{equation}
\hat{\chi}_j(q) \equiv e^{-i\pi(h_j-c/24)}\;\chi_j(-\sqrt{q})
\end{equation}
The modular transformation to be applied here is then given by the matrix
\begin{equation}
P = T^{1/2} S T^2 S T^{1/2}.
\end{equation}
The calculation of this matrix was performed in \cite{SS} and can be
found in the appendix of \cite{BCW}. It is shown there that
\begin{equation}
\label{P}
P_l{}^j = \frac{2}{\sqrt{k+2}}\;
        \sin\left(\frac{\pi(2l+1)(2j+1)}{2(k+2)}\right) E_{2l+2j+k}\ .
\end{equation}

The M\"obius amplitude for $n_r$ spherical branes of type $r$ reads
\begin{equation}
\M^r = n_r \eps'_r \sum_{\substack{l=0 \\ l\rm\ integer}}^{2r}\;
             ( (-)^l \hat{\chi}_l(q) + \eps''_r\; \hat{\chi}_{k/2-l}(q) )
\end{equation}
where $\eps'_r$ is a global sign, as always in M\"obius amplitudes,
the two terms correspond to the two sorts of strings mentioned in the
previous subsection, and $\eps''_r$ is for the projection of strings
of the second type.

Then, after a modular transformation with the $P$ matrix, the
consistency condition
\begin{equation}
\M^r = n_r \sum_j C^j D^j_r \hat{\chi}_j(\tilde{q})
\end{equation}
implies that $\eps_j$ is a constant $\eps$, $\eps'_r = (-)^{2r} \eps$,
and $\eps''_r = (-)^{2r} \zeta$.

As the photons are in the isospin zero representation of the Kac-Moody
algebra, the gauge group is given by the sign before $\chi_0$, i.e.\
$\eps'_r$. So, as for the $\O0$ of the 3-sphere, one has an alternance
of orthogonal and symplectic groups.

For the equatorial branes, one has to distinguish between whether
$\gamma$ commutes or anticommutes with $Z$. In the first case, the
states that appear in the M\"obius amplitude are $|++\>$ and $|--\>$,
which are of even isospin, so the M\"obius amplitude reads
\begin{equation}
\begin{array}{lll}
\M^R & = \displaystyle (n_+ + n_-) \,\eps'_R
           \sum_{\substack{l=0 \\ l\rm\ even}}^{k/2} \hat{\chi}_l(q) \\
     & = \displaystyle \frac{1}{2}(n_+ + n_-) \,\eps'_R
           \sum_{\substack{l=0 \\ l\rm\ integer}}^{k/2} 
           ((-)^l \hat{\chi}_l(q) + \hat{\chi}_{k/2-l}(q))\ .
\end{array}
\end{equation}
In the second case, the states that appear in the M\"obius amplitude
are $|+-\>$ and $|-+\>$, which are of odd isospin, so the M\"obius
amplitude reads
\begin{equation}
\begin{array}{lll}
\M^R & = \displaystyle 2n\,\eps'_R
           \sum_{\substack{l=0 \\ l\rm\ odd}}^{k/2} \hat{\chi}_l(q) \\
     & = \displaystyle -n\,\eps'_R
           \sum_{\substack{l=0 \\ l\rm\ integer}}^{k/2}
           ((-)^l \hat{\chi}_l(q) - \hat{\chi}_{k/2-l}(q))\ .
\end{array}
\end{equation}
In both cases, up to a factor $1/2$, which is also present in the
couplings $D^j_R$ \eqref{DjR}, this amplitude has the same form as the
other amplitudes with $\eps''_R$ equal to $+1$ in the first case, and
$-1$ in the second case. The consistency condition is then the same as
for spherical branes, i.e.\ $\eps'_R = \eps''_R\,(-)^{k/2} \eps$ and
$\zeta = \eps''_R\,(-)^{k/2}$. Thus:
\begin{itemize}
\item If $\zeta = (-)^{k/2}$, then the gauge group is $SO(n_+) \times
SO(n_-)$ or $USp(n_+) \times USp(n_-)$, depending on $\eps_R$.
\item If $\zeta = -(-)^{k/2}$, then one must have $n_+ = n_- \equiv n$,
and the gauge group is $U(n)$.
\end{itemize}

\acknowledgments 
I thank C.~Bachas, P.~Windey, P. Bordalo and V. Schomerus for useful
conversations.


\begin{thebibliography}{999}
\bibitem{GW}
D.~Gepner and E.~Witten, {\it String theory on group manifolds},
\npb{278}{1986}{493}

\bibitem{Car}
J.L.~Cardy,
{\it Boundary conditions, fusion rules and the Verlinde formula}, 
\npb{324}{1989}{581}

\bibitem{Sag}
G.~Pradisi, A.~Sagnotti and Y.S.~Stanev,
{\it Planar duality in $SU(2)$ WZW models},
\plb{354}{1995}{279} [\hepth{9503207}];
{\it The open descendants of nondiagonal $SU(2)$ WZW models},
\plb{356}{1995}{230} [\hepth{9506014}];
{\it Completeness conditions for boundary operators in 2D conformal
field theory}, \plb{381}{1996}{97} [\hepth{9603097}].

\bibitem{SS}
A.~Sagnotti and Y.~S.~Stanev,
{\it Open Descendants in Conformal Field Theory},
\forp{44}{1996}{585-596} [\hepth{9605042}].

\bibitem{AS}
A.Y.~Alekseev and V.~Schomerus, {\it D-branes in the WZW model},
\prd{60}{1999}{061901} [\hepth{9812193}].

\bibitem{FFFS}
G.~Felder, J.~Fr\"ohlich, J.~Fuchs and C.~Schweigert,
{\it The geometry of WZW branes},
\jgp{34}{2000}{162-190} [\hepth{9909030}].

\bibitem{DMS}
E.~Dudas, J.~Mourad and A.~Sagnotti,
{\it Charged and Uncharged D-branes in various String Theories},
\npb{620}{2002}{109} [\hepth{0107081}].

\bibitem{MSS}
K.~Matsubara, V.~Schomerus and M.~Smedb\"ack,
{\it Open Strings in Simple Current Orbifolds}, \hepth{0108126}.

\bibitem{BDS}
C.~Bachas, M.~Douglas and C.~Schweigert,
{\it Flux stabilization of D-branes}, 
\jhep{05}{2000}{048} [\hepth{0003037}].

\bibitem{BRS}
P.~Bordalo, S.~Ribault and C.~Schweigert,
{\it Flux stabilization in compact groups},
\jhep{10}{2001}{036} [\hepth{0108201}].

\bibitem{GP}
E.~Gimon and J.~Polchinski,
{\it Consistency Conditions for Orientifolds and D-Manifolds},
\prd{54}{1996}{1667} [\hepth{9601038}].

\bibitem{Bru}
I.~Brunner,
{\it On orientifolds of WZW models and their relation to geometry},
\jhep{01}{2002}{007} [\hepth{0110219}].

\bibitem{HSS}
L.R.~Huiszoon, K.~Schalm and A.N.~Schellekens,
{\it Geometry of WZW Orientifolds}, \hepth{0110267}.

\bibitem{BCW}
C.~Bachas, N.~Couchoud and P.~Windey,
{\it Orientifolds of the 3-sphere},
\jhep{12}{2001}{003} [\hepth{0111002}].

\bibitem{ARS}
A.Y.~Alekseev, A.~Recknagel and V.~Schomerus, 
{\it Brane dynamics in background fluxes and non-commutative geometry},
\jhep{05}{2000}{010} [\hepth{0003187}].

\end{thebibliography}
\end{document}